\documentclass[iop]{emulateapj}
\usepackage{amsmath}
\usepackage{epstopdf}
\usepackage{multirow}
\usepackage{subfigure}
\usepackage{longtable}
\newcommand{\mearth}{M$_{\oplus}$}
\newcommand{\rearth}{R$_{\oplus}$}
\newcommand{\msun}{M$_{\odot}$}

\newcommand{\mpers}{m~s$^{-1}$}
\newcommand{\teff}{$T_{\rm eff}$}

\slugcomment{Accepted to {\it The Astrophysical Journal} 13 May 2013}
\shorttitle{Giant Planets around Late K Dwarf Stars}

\shortauthors{Gaidos et al.}

\begin{document}


\title{An Understanding of the Shoulder of Giants: Jovian Planets around Late K Dwarf Stars and the Trend with Stellar Mass}

\author{Eric Gaidos} 
\affil{Department of Geology \& Geophysics, University of Hawai`i at M\={a}noa, Honolulu, HI 96822}
\email{gaidos@hawaii.edu}

\author{Debra A. Fischer}
\affil{Department of Astronomy, Yale University, New Haven, CT 06520}

\author{Andrew W. Mann}
\affil{Institute for Astronomy, University of Hawai`i at M\={a}noa, Honolulu, HI 96822}

\author{Andrew W. Howard}
\affil{Institute for Astronomy, University of Hawai`i at M\={a}noa, Honolulu, HI 96822}

\begin{abstract}
  Analyses of exoplanet statistics suggest a trend of giant planet
  occurrence with host star mass, a clue to how planets like Jupiter
  form.  One missing piece of the puzzle is the occurrence around late
  K dwarf stars (masses of $0.5-0.75$\msun{} and effective
  temperatures of 3900-4800~K).  We analyzed four years of Doppler
  radial velocities data of 110 late K dwarfs, one of which hosts two
  previously reported giant planets.  We estimate that $4.0\pm2.3$\%
  of these stars have Saturn-mass or larger planets with orbital
  periods $< 245$~d, depending on the planet mass distribution and RV
  variability of stars without giant planets.  We also estimate that
  $0.7\pm0.5$\% of similar stars observed by \emph{Kepler} have giant
  planets.  This \emph{Kepler} rate is significantly (99\% confidence)
  lower than that derived from our Doppler survey, but the difference
  vanishes if only the single Doppler system (HIP~57274) with
  completely resolved orbits is considered.  The difference could also
  be explained by the exclusion of close binaries (without giant
  planets) from the Doppler but not \emph{Kepler} surveys, the effect
  of long-period companions and stellar noise on the Doppler data, or
  an intrinsic difference between the two populations.  Our estimates
  for late K dwarfs bridge those for solar-type stars and M dwarfs and
  support a positive trend with stellar mass.  Small sample size
  precludes statements about finer structure, e.g. a ``shoulder'' in
  the distribution of giant planets with stellar mass.  Future surveys
  such as the Next Generation Transit Survey and the Transiting
  Exoplanet Satellite Survey will ameliorate this deficiency.
\end{abstract}

\section{Introduction}

Parent star mass is a fundamental parameter in planet discovery space
because there are both theoretical predictions and observational
evidence that the properties of planetary systems depends on central
star mass.  In addition, it is of practical significance: the
sensitivity of the Doppler, astrometric, and transit techniques of
planet detection scale inversely with stellar mass (or radius), such
that smaller planets can be detected around less massive stars (or
smaller) stars.

Most surveys for planets, i.e., ground-based radial velocity (RV)
surveys and the \emph{Kepler} transit survey mission, are flux- or
magnitude-limited at visible wavelengths, favoring the inclusion of
intrinsically bright stars.  In addition, the Doppler method only
works with spectra having large numbers of deep absorption lines,
which disqualifies hot stars, and the transit method works better for
cool dwarfs, around which planets produce deeper transits.  These
opposing trends mean that catalogs of exoplanet-hosting stars are
dominated by solar-type stars with late F to early K spectral types
\citep{Udry2007_PPV,Batalha2010}.  Moreover, a focus on solar-mass stars
satisfies a desire to determine the occurence, nature, and potential
habitability of planets around other stars like the Sun.

On the other hand, M dwarf stars are now widely recognized as an
attractive ``short cut'' to the discovery of Earth-like planets
because such stars are numerous, small, and their habitable zones are
close-in, meaning that planets orbiting within them will be more
detectable by the Doppler or transit methods.  Doppler surveys have
included the few, nearby M dwarfs that are sufficiently bright at
visible wavelengths \citep{Bonfils2013}, and high-precision infrared
spectrographs are being constructed to take advantage of the greater
emission of these stars at longer wavelengths
\citep{Artigau2011,Quirrenbach2012}.  Several thousand M dwarfs were
added to the \emph{Kepler} catalog for these reasons
\citep{Batalha2010}.

Between the early K-type dwarfs and M dwarfs are the late K dwarfs,
having K4-K7 spectral subtypes, $T_{\rm eff} \approx 3900-4800$~K, and
$M_* \approx 0.5-0.75$\msun{}.  These stars have been comparatively
neglected in planet surveys because they are intrinsically faint and
because they are not M dwarfs.  The two largest Doppler surveys, the
California Planet Search (CPS) and the HARPS survey, include
relatively few late K stars.  Ironically, these stars may be
especially attractive targets for Doppler surveys because intrinsic
stellar Doppler noise or ``jitter'' decreases with later spectral type
and could be $<1$~m~s$^{-1}$ among K dwarfs
\citep{Isaacson2010,Lovis2011}.  The K5 dwarf HD~85512 is one of the
most Doppler-stable stars reported (residual RMS = 0.75~m~s$^{-1}$), a
property that has permitted the discovery of a super-Earth near or
inside its habitable zone \citep{Pepe2011}.

Giant planets, defined here as planets with mass greater than that of
Saturn (95\mearth{}) or radius greater than 8\rearth{}, are readily
detected by Doppler observations with a sufficient time baseline if
the planets orbit within $\sim 1$~AU of their host stars.  Their
distribution with mass or spectral type of the host star can test the
core accretion scenario of giant planet formation as well as models of
orbital migration.  Numerous studies have found evidence that the
fraction of stars with giant planets increases with stellar mass
\citep{Fischer2005b,Cumming2008,Johnson2010}.  Likewise,
\citet{Fressin2013} estimated that the occurrence of giant planets is
lower for \emph{Kepler} M dwarfs than for G and K dwarfs.  These
findings support a theoretical prejudice that a more massive star is
born with a more massive disk that can spawn the solid cores capable
of accreting disk gas before its dispersal \citep{Laughlin2004}.
However, any relation between disk mass and stellar mass is ambiguous
\citep{Williams2011}.  Moreover, the picture below $M_* =
0.75$\msun{}~is unclear because the statistics are poor; the sample of
\citet{Johnson2010} included only 142 late K and M dwarfs with 5
reported giant planets.  Likewise, the difference in giant planet
occurrence around solar-type and M dwarf stars reported by
\citet{Fressin2013} has only a 1.3$\sigma$ significance.  K dwarfs
provide the ``missing link'' in this picture, and surveys for giant
planets could reveal whether the difference in giant planet frequency
exists, and whether it is a smooth transition or an abrupt
``shoulder''.

The M2K Doppler survey targets the brightest late K dwarfs, bridging
the gap between solar-type stars and M dwarfs.  The survey reported
one giant planet around an M3 dwarf \citep{Apps2010} and a triple
system including two Saturn- to Jupiter-mass planets around the K4-K5
dwarf HIP~57274 (GJ~439) with $T_{\rm eff} \approx 4640\pm100$K and $M_*
\approx 0.73\pm0.05$, based on Yale-Yonsei isochrones
\citep{Fischer2012}.  Only four other mid- to late K-type dwarf hosts
of giant planets have been reported: Jupiter-mass HAT-20b transits a
similar K3 star with $T_{\rm eff} \approx 4619 \pm 72$K
\citep{Bakos2011,Torres2012}.  The effective temperatures of the other
three late K-type hosts (WASP-43b, HIP~70849, and WASP-80b) are not
well established (Table \ref{tab.planets} and
Sec. \ref{subsec.future}).  All other K dwarf hosts are hotter and
have earlier spectral subtypes.  We use the results of the M2K survey
and {\it Kepler} to establish new constraints on the occurrence of
giant planets around late K dwarfs and compare them with values for
solar-type stars and M dwarfs.

\section{Methods: M2K Survey}

\subsection{Sample, Observations, and Reduction}

\begin{figure}
\epsscale{0.8}
\plotone{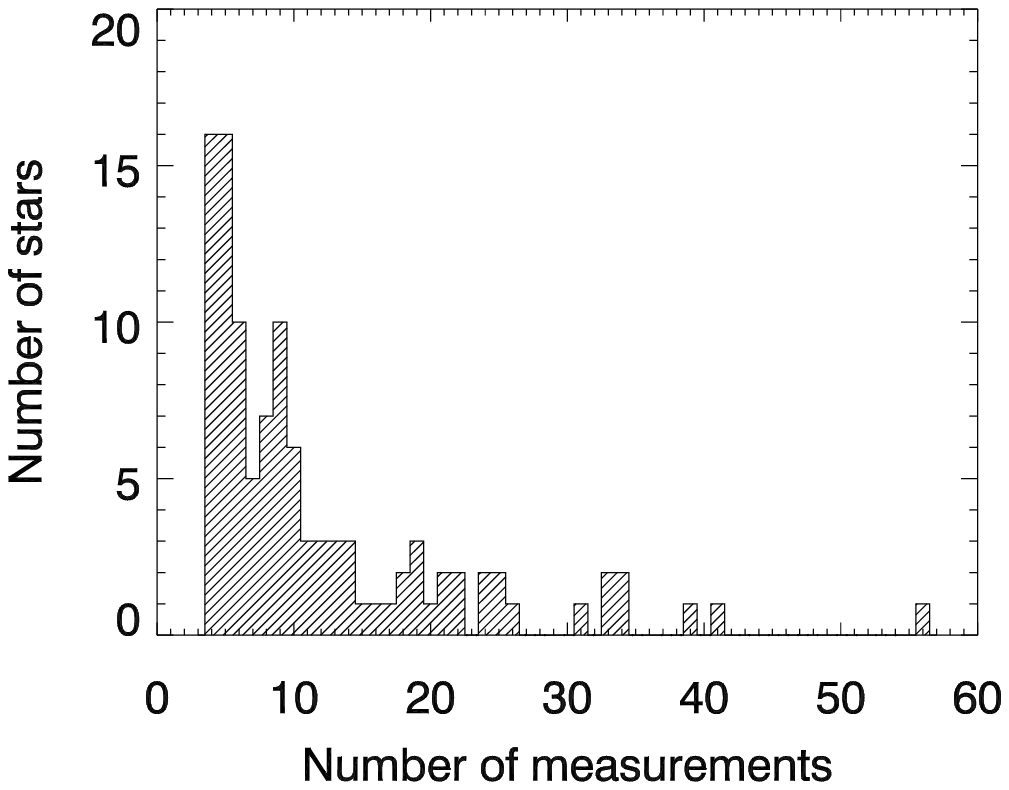}
\caption{Distribution of number of Doppler measurements per star for
  those M2K stars selected for the analysis of giant planet fraction.
  One star (HIP~57274) with two giant planets has 120 observations and
  is off scale. \label{fig.nobshist}}
\end{figure}

Doppler observations were performed with the High Resolution Echelle
Spectrograph (HIRES) on the Keck I telescope \citep{Vogt1994}.
Observations obtained $R=55,000$ with the B5 decker and a typical SNR
of 200.  Wavelength calibration was provided by a molecular iodine
cell in the beamline.  Radial velocity solutions were obtained by a
forward modeling process in which an intrinsic stellar spectrum is
obtained without the iodine cell, multiplied by a R $\sim 5 \times
10^5$ spectrum of the iodine cell taken with a Fourier Transform
Spectrograph, and convolved by the instrumental profile.  The relative
shift in wavelength between the model and observed spectra is a free
parameter.  The median formal measurement error in the M2K survey is
1.25~\mpers{}.  We obtained $N=4$ or more RV measurements on 159
stars.  The median number of measurements for each star used in our
analysis is $N=9$, however the distribution of number of measurements
is very uneven because of our strategy of follow-up of RV-variable
stars (Fig. \ref{fig.nobshist}).

\begin{figure}
\epsscale{0.8}
\plotone{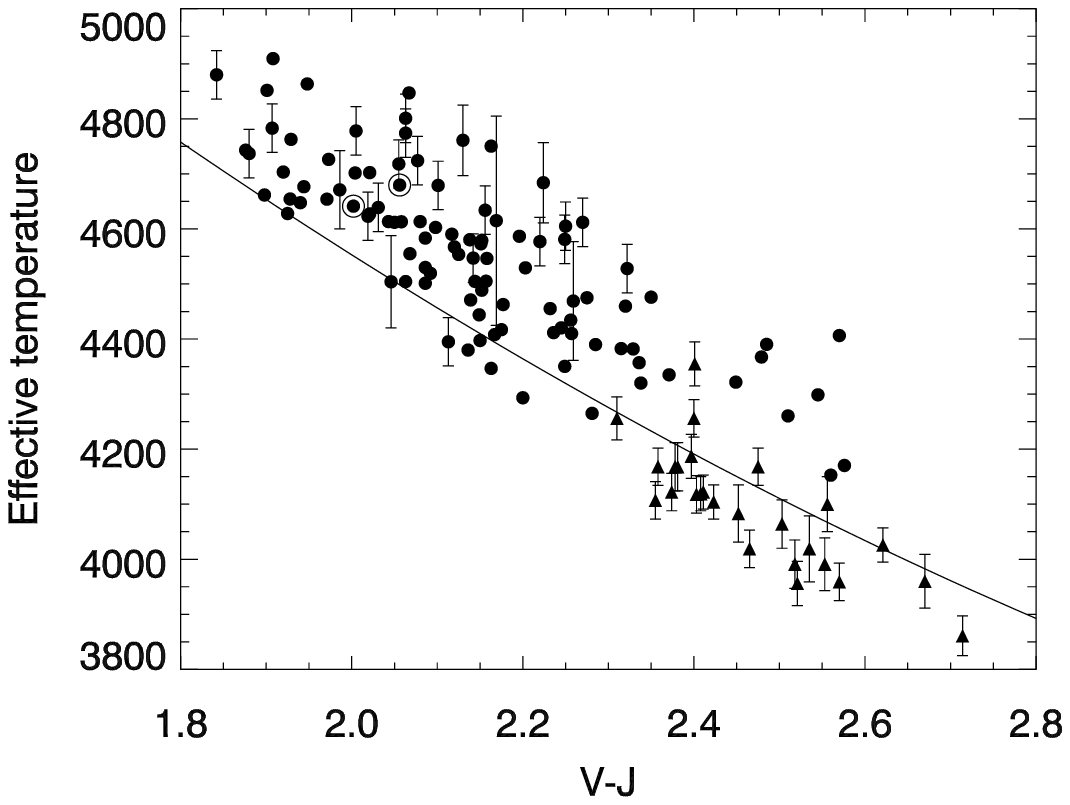}
\caption{Stellar effective temperature vs. $V$-$J$ color of late K
  dwarfs in the M2K Doppler survey.  Circles represent temperatures
  from SME analyses of high-resolution spectra \citep{Valenti1996},
  whereas triangles represent temperatures from fitting
  medium-resolution spectra to PHOENIX synthetic spectra (Mann et al.,
  in prep) and calibrating on stars in \citet{Boyajian2012}.  Only
  some error bars are shown for clarity.  The solid curve is an
  empirical \teff{} vs. $V$-$J$ relation from \citet{Boyajian2012}.
  Two systems with published giant planets (HIP~57274 and HIP~2247)
  are circled. \label{fig.temp_vs_vj}}
\end{figure}

Ideally we would have defined our sample of late K dwarfs in terms of
effective temperature \teff{}, a fundamental stellar parameter, but
not all our stars have spectroscopically-determined values.  Thus we
selected stars based on $V-J$ color as a proxy for \teff{}.  We
restricted the sample to 140 stars with $1.8 < V-J < 2.8$,
corresponding to $3900 < T_{\rm eff} < 4750$~K or spectral subtypes
K3-K7 \citep{Gray2009}, based on an empirical color-\teff{} relation
using stars with measured angular diameters \citep{Boyajian2012}.  We
corroborated this selection by estimating the \teff{} of many of these
stars using spectra (Fig. \ref{fig.temp_vs_vj}).  Stellar parameters
(including \teff{} and metallicity [Fe/H]) of stars with $1.8 < V-J <
2.3$,were estimated using the Spectroscopy Made Easy package
\citep{Valenti1996} (Table 2).  SME performs poorly on dwarfs with
$V-J > 2.3$ and $T_{\rm eff}\le 4300$~K so for these stars estimated
\teff{} by comparing moderate-resolution ($R \sim 1500$)
visible-wavelength (3500-8500\AA) spectra obtained with the Supernova
Integral Field Spectrograph \citep{Lantz2004} on the University of
Hawaii 2.2~m telescope with synthetic spectra generated by the PHOENIX
BT-SETTL program \citep{Allard2011}.  The comparison procedure is
described in \citet{Lepine2013} and was adjusted (Mann et al., in
prep.) to maximize agreement with the calibrator stars of
\citet{Boyajian2012}.  For stars without spectra, we estimated \teff{}
using $V-J$ color and the \citet{Boyajian2012} relation (Table 2).

The median offset between SME-derived and $V-J$-based temperatures is
140~K.  For this reason, we included the 13 stars with SME-based
$T_{\rm eff} > 4750$~K (but acceptable $V-J$) as their actual effective
temperatures are probably within the acceptable range.
Unsurprisingly, \teff{} estimates based on calibrated
comparisons between moderate-resolution spectra and PHOENIX models are
consistent with the \citet{Boyajian2012} relation.  Nearly all of our
stars have measured parallaxes and we estimated masses using the
relation with absolute $K$ magnitude in \citet{Henry1993}.  For those
few stars lacking parallaxes we used the empirical relations between
\teff{}, stellar radius, and stellar mass in
\citet{Boyajian2012}.

\begin{figure}
\epsscale{0.8}
\plotone{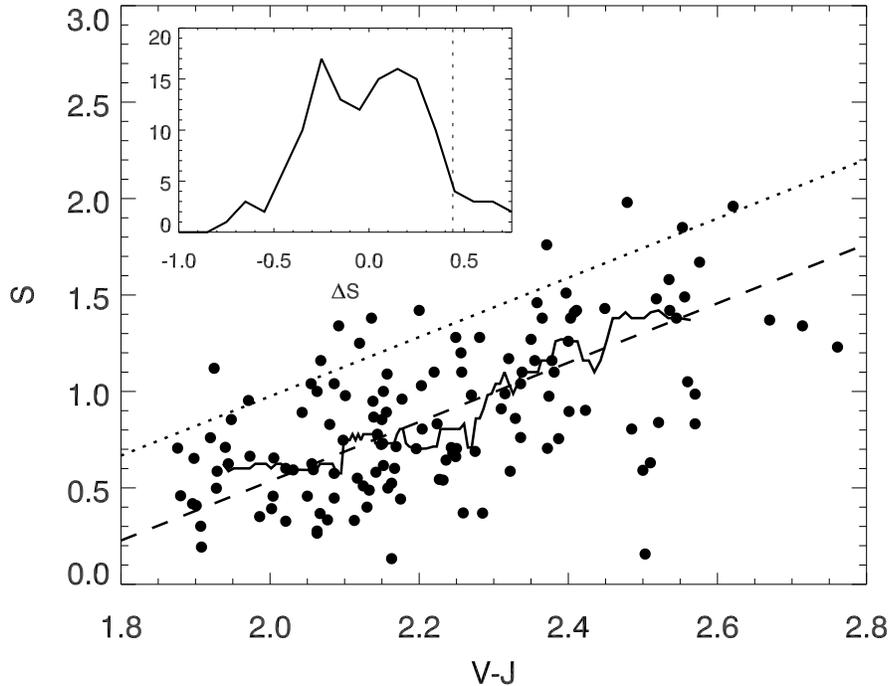}
\caption{Flux $S$ in the core of the Ca II HK lines, normalized by the
  continuum, vs. $V$-$J$ color.  High values of $S$ are associated
  with elevated stellar activity and astrophysical Doppler noise or
  ``jitter''.  The solid line is a running median ($N=20$), the dashed
  line is a linear regression of the median $\bar{S}$, and the dotted
  line is the linear fit + 0.44, above which stars were excluded from
  the analyis.  This threshold was selected based on the distribution
  of $\Delta S = S - \bar{S}$ (inset).  Eleven stars or 8\% of the
  sample were excluded based on this criterion. }
\label{fig.s_vs_vj}
\end{figure}

We excluded stars exhibiting relatively high emission in the H and K
lines of Ca II.  These stars are chromospherically active and tend to
exhibit higher astrophysical Doppler noise or ``jitter''
\citep[e.g. ][]{Isaacson2010}.  Figure \ref{fig.s_vs_vj} shows values of
$S_{HK}$, the flux in the Ca II line cores normalized by the
continuum, vs. $V-J$ color.  The trend of increasing $S_{HK}$ with
redder $V-J$ is due to the lower blue continuum rather than elevated
Ca II emission, in redder stars.  We calculated a running median
($N=20$), fit a linear function $\bar{S}_{HK}$ with $V-J$, and
subtracted that from these values.  The histogram (inset of
Fig. \ref{fig.s_vs_vj}) suggests a cutoff at $S_{HK} - \bar{S}_{HK} =
0.44$, which rejects 11 stars as exceptionally active.  Among the
stars we admitted were 5 for which it was not possible to estimate
$S_{HK}$ because the continuum was not detected.

We next considered the distribution of RV standard deviations (RMS) of
the remaining 129 stars.  The majority (100) of the stars fall in a
cluster with RMS $<15$~\mpers{} (Fig. \ref{fig.rmshist}).  We
inspected the RV data of the 29 stars with RMS $>15$~\mpers{}.  One of
these is HIP~57274, which has been described elsewhere
\citep{Fischer2012}.  Eight others have additional stars within 5 arc
sec and were excluded because leakage of light into the spectrograph
slit is an established source of RV error.

We analyzed each of the remaining 20 sets of RVs with both weighted
linear and quadratic regressions and applied an F-test to evaluate the
significance of any reduction in variation after subtraction of the
fit.  Although all stars have $\ge 4$ RV measurements, and thus the
number of degrees of freedom for a quadratic fit is $\ge1$, irregular
sampling means that over-fitting and erroneous reduction in RV
variation is possible.  For example, four observations in two very
closely-spaced pairs cannot be reliably regressed: a significant
radial velocity offset between the pairs could be the result of a
linear trend or any RV variation.  To identify such cases, we computed
an effective $N$ equal to the sum of normalized Voronoi-type weights
$w_i = (t_{i+1}-t_{i-1})/2$ that are often assigned by regularization
algorithms \citep{Strohmer2000}.  End points have weights $w_1 = t_2 -
t_1$ and $w_N = t_N-t_{n-1}$.  Thus
\begin{equation}
N_{\rm eff} = \frac{3(t_N-t_1) - (t_{N-1}-t_2)}{2\; {\rm max}(w_i)}.
\end{equation}
In the limit of large $N$, $N_{\rm eff}$ and thus the maximum order of
the polynomial that should be used in a regression approach the total
time interval divided by the maximum interval between points.  This is
analogous to the Nyquist sampling criterion.

Doppler datasets with RMS $>$15~\mpers{} were processed in one of
three ways: (i) If regressions did not significantly reduce variance
(F-test $p < 0.05$) the data were analyzed as is. (ii) If a regression
did significantly reduce variance and $N_{\rm eff}$ was $>3$ (or $>4$,
in the case of a quadratic), the best fit is subtracted before
analysis. (iii) If a regression is significant but $N_{\rm eff}$ is
not sufficiently large, the star and its data were excluded from the
analysis.  We excluded 11 stars in this way, leaving 110 stars for
analysis, including HIP~57274.

\begin{figure}
\epsscale{0.8}
\plotone{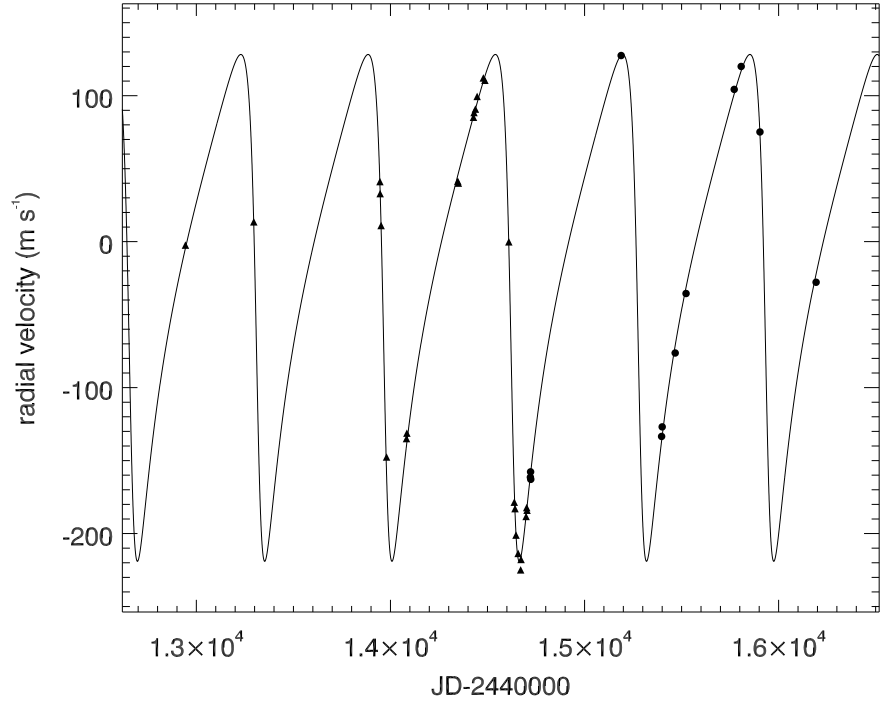}
\caption{Thirty-eight radial velocities of HIP~2247 showing barycenter
  motion produced by the long-period giant planet discovered by
  \citet{Moutou2009}\label{fig.hip2247}.  Triangles are Moutou et
  al. measurements with HARPS and circles are M2K measurements with
  Keck-HIRES.  The solid line is the best-fit Keplerian orbit with $m
  \sin i = 5.14$~$M_J$, $P_K=655.9$~d, and $e=0.543$.}
\end{figure}

HIP~2247 has a long-period super-Jupiter previously identified by
\citet{Moutou2009}.  We fit the combined HARPS and Keck-HIRES data
using the RVLIN code \citep{Wright2009}, generating errors using 100
Monte Carlo realizations of the data by randomly reshuffling the
residuals to the previous fit.  We find essentially the same planetary
parameters as \citet{Moutou2009}, but with significantly reduced
uncertainties: $M_p \sin i = 5.14\pm0.02 M_J$, $P=655.90\pm0.22$~d, and
$e = 0.543 \pm 0.0011$, with a residual RMS of 3.8~\mpers{}
(Fig. \ref{fig.hip2247}).  No significant trend was found ($-0.0024
\pm 0.0011$~\mpers{}).  The uncertainty in $m \sin i$ does not include
errors in the estimated stellar mass of 0.77\msun{}.  Any giant
planets with $P < 245$~d can be ruled out: we include this system in
our sample but consider it a definitive non-detection.

\begin{figure}
\epsscale{0.8}
\plotone{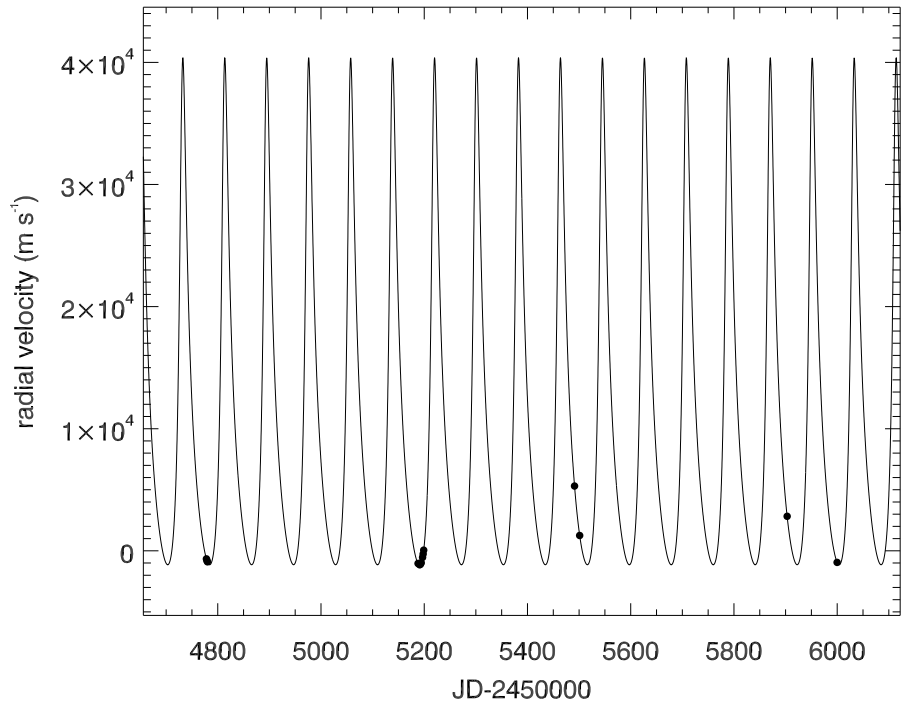}
\caption{Fifteen M2K radial velocities of the the K+M binary star
  system HIP~38117.  The best-fit Keplerian orbit has $P_K=81.28$~d
  and $e=0.478$ \label{fig.hip38117}}
\end{figure}

HIP~38117 exhibits RV variation consistent with the presence of a
stellar-mass companion on a $81.28\pm0.01$~d orbit with an
eccentricity of $0.478\pm0.012$ (Fig. \ref{fig.hip38117}).  Assuming a
primary mass of 0.73\msun{} based on the system's $V-J$ color, and
assuming that the secondary contributes negligible flux, the
companion's $M_* \sin i$ is $0.45\pm0.16$\msun{}, i.e. this is a very
late K or M dwarf.  (This calculation assumes an average value of
$\langle \sin i \rangle = \pi/4$ to calculate the total system mass.)
The residual RMS is 3~m~s$^{-1}$ ($N=15$).  As a planetary orbit with
a comparable orbital period is unlikely to be stable, we follow the
suit of other studies by excluding this binary system from our
analysis.

\subsection{Estimation of Planet Fraction \label{subsec.estimate}}

We estimated the fraction of stars $f$ with giant planets having $M_P
> 0.3M_J$ (i.e. Saturn mass) and Keplerian orbital periods 1.7~d$ <
P_K <245$~d.  The choice of outer cutoff in $P_K$ is based on the
temporal baseline of our data - nearly all stars were monitored for at
least 245~d - and motivated by the longest bin with good statistics in
the analysis of \emph{Kepler} planet candidates by \citet{Fressin2013}
.  The inner cutoff corresponds to the location of the rollover in the
period distribution of giant planets around \emph{Kepler} stars
\citep{Howard2012}.  We construct and maximize a likelihood function
to find the most probable value of $f$ and its uncertainty.  The
details of the calcuations are given in the Appendix and the method is
only summarized here.

A standard procedure to estimate the fraction of stars with planets is
to maximize a binomial expression involving the product of detections
and non-detections.  However, with RV data it can be difficult or
impossible to rule out all possible planets, e.g. those on face-on
orbits.  Thus we replace detections and non-detections wih a Bayesian
statistic that is sum of the probabilities $p_i^0$ and $p_i^1$ that
there are zero or one giant planets around the $i$th star, with $1-f$
and $f$ as \emph{priors} for zero or one planets, respectively:
\begin{equation}
\label{eqn.likely}
\ln \mathcal{L} = \displaystyle\sum_i \ln\left(p_i^0(1-f) + p_i^1f \right),
\end{equation}
where the sum is over all stars.  This counts multi-giant planet
systems once, and thus underestimates the planet \emph{occurrence}
(planets per star).  The probabilities $p_i^0$ and $p_i^1$ are
Gaussian functions of the difference between the predicted and
observed radial velocities $v_{ij}$ and $\hat{v_{ij}}$, weighted by
priors $\tilde{p_i}$ marginalized over all model parameters
\begin{equation}
\label{eqn.prob}
p_i^n = \langle \tilde{p_i}\exp \left[-\displaystyle\sum_j \frac{\left(v_{ij}-\hat{v}_{ij}^n\right)^2}{2\sigma_{ij}^2}\right] \rangle,
\end{equation}
and where $\sigma_{ij}$ are the formal errors, astrophysical noise or
``jitter'', as well as systematic error and the contribution of small
planets to motion of the star around the system's barycenter (see
Sec. \ref{subsec.results}), added in quadrature.  This method is
analogous to the approach used in \citet{Gaidos2012} but uses the
individual RV measurements, not the RMS.

The Gaussian form of Eqn. \ref{eqn.prob} means that only the best-fit
sets of parameter values (barycenter motion $v_0$, Keplerian period
$P_K$, Doppler amplitude $K$, eccentricity $e$, longitude of
periastron $\omega$ and time of periastron $t_0$) make significant
contributions to $p^0$ or $p^1$.  We used the linear dependence of the
radial velocities on the barycenter velocity $v_0$ and Doppler
amplitude $K$ to analytically solve for the best-fit values of these
two parameters given values for the other parameters.  To marginalize
over planet mass we used the relation between $K$, planet mass, and
orbital inclination and adopt a power-law distribution of log mass
with an index $\alpha = -0.31$ \citep{Cumming2008}.  (The sensitivity
of our results to this value is explored in Section
\ref{subsec.sensitivity}.)  We marginalized Eqn. \ref{eqn.prob} over
the full range of possible values of $e$, $\omega$, $t_0$, using a
Rayleigh function for the prior on eccentricity \citep{Moorhead2011a}
and uniform priors for $\omega$ and $t_0$.  We further evaluated the
probability over $3\;{\rm d} < P_K < 245$~d at intervals of equal
prior probability, assuming a power-law distribution for log period
having index $\beta = 0.26$ \citep{Cumming2008}.  To better sample
intervals of $P_K$ corresponding to higher probablity we used each
$P_K$ value as an initial value in a fit of a Keplerian solution to
the data with the RVLIN routine \citep{Wright2009}.  We used the
fixed, best-fit values of the other parameters for the RVLIN fit,
obtained an adjusted value of $P_K$, and then re-calculated the other
parameters as described above.  This procedure was repeated two times,
which we found was sufficient for convergence.

We calculated $p^0$ and $p^1$ by summing over final values of all the
parameters, and normalizing by $p^0 + p^1$.  For three stars $p_0$ and
$p_1$ were both incalculably small due to large disagreements between
$v$ and $\hat{v}$ for either the zero- or one-planet models.  This
could be due to elevated stellar jitter or the presence of smaller
planets (see below).  For these stars we assigned $p_0 = p_1 = 0.5$,
i.e. the zero- and one-planet models are accepted or rejected with
equal likelihood.  We then evaluated Eqn. \ref{eqn.likely} over all
possible values of $f$, and found the maximum.  We calculated an
approximate uncertainty by assuming asymptotic normality, iteratively
fitting a parabola to the log-likelihood curve, and assigning
$\sigma_f = 1/\sqrt{2C}$, where $C$ is the curvature coefficient of
the parabola.

\begin{figure}
\epsscale{0.78}
\plotone{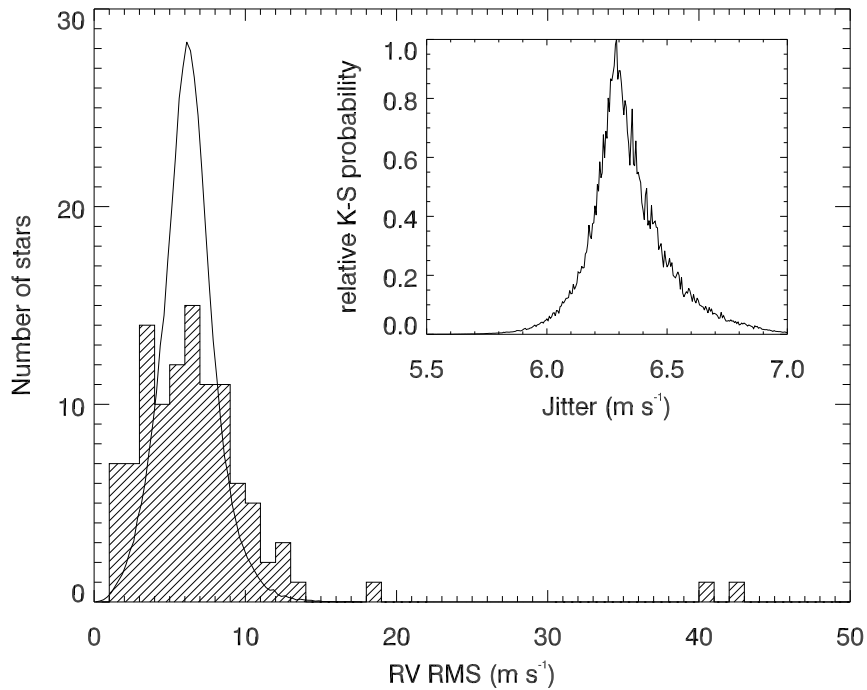}
\caption{Distribution of adjusted radial velocity RMS among 110 M2K
  stars after excluding or adjusting case of high RMS.  One star with
  RMS = 65~\mpers{} falls outside the plot.  Systems with RMS
  $>15$~\mpers{} were either excluded or significant linear/quadratic
  trends fitted and removed (see text).  The solid curve is the
  best-fit model for the resulting distribution at RMS $<15$~\mpers{}
  assuming pure Gaussian-distributed noise that is the sum of formal
  errors and an astrophysical noise term $\sigma_0$ that includes both
  stellar jitter and barycenter motion due to small planets.  The
  value $\sigma_0 = 6.3$~\mpers{} which best reproduces the observed
  distribution was selected by maximizing the Kolmogorov-Smirnov
  statistic that the actual and model RMS values are drawn from the
  same distribution (inset).}
\label{fig.rmshist}
\end{figure}

Both the zero- and one-planet models do not account for barycenter
motion due to the presence of other, smaller planets, as well as any
sources of systematic error.  As a result, for some stars both models
are strongly rejected, leading to an erroneously high value of $f$.
To account for this effect, we treat this barycenter motion as an
additional source of uncorrelated, random RV noise or ``jitter'' that,
along with stellar noise, can be described by a single value of
$\sigma_0$.  A value for $\sigma_0$ was chosen by \emph{assuming} that
the pronounced cluster of systems with RMS $<15$~\mpers{}
(Fig. \ref{fig.rmshist}) represents stars without giant planets, and
fitting that distribution by a Monte Carlo model.  We constructed 1000
artificial realizations of the data with the same number of RVs per
star but drawn from a random normal distribution.  The variance of
this distribution was set equal to the formal measurent error and a
trial value of $\sigma_0$ added in quadrature.  We computed the RMS
values and comparing the distribution to the observed distribution
(after subtracting any trends) with a two-sided Kolmogorov-Smirnov
test.  We found that a curve with $\sigma_0 = 6.3$~\mpers{} (solid
curve in Fig. \ref{fig.rmshist}) maximizes the K-S probability of the
Monte Carlo distribution (inset of Fig. \ref{fig.rmshist}), and we use
that value in our estimation of $f$.  The poor agreement between the
observed and ``best-fit'' distributions reflects the inability of a
single value of $\sigma_0$ to capture the diversity of stellar RV
behavior.

\section{Methods: \emph{Kepler} Survey}

We compared our estimate of the fraction of M2K dwarfs with giant
planets with one for late K dwarfs observed by \emph{Kepler}.  We
selected \emph{Kepler} targets with $1.8 < V-J < 2.8$, with $V$
magnitudes estimated using the relation $V = r + 0.44(g-r) - 0.02$
\citep{Fukugita1996}.  We further limited the sample to stars that had
been observed in at least seven of quarters Q1-Q8.  The absence of a
single quarter will minimally affect the detection efficiency but is
common because some stars were added after Q1 and others fall within
\emph{Kepler}'s defunct CCD module during one of four rotations of the
spacecraft.  Stellar and planetary parameters of \emph{Kepler} stars
were estimated by fits to Dartmouth stellar models \citep{Dotter2008}
using the Bayesian procedure described in \citet{Gaidos2013b}.  We
restricted the analysis to 6293 dwarf stars with $3900 < T_{\rm eff} <
4800$~K, $\log g > 4$, and $K_P < 16$.  In this sample are two giant
planet candidates with $P_K < 245$~d: KOI~1176.01 is a hot Jupiter
($P_K = 1.94$~d) orbiting a star with $T_{\rm eff} \approx 4625$~K.  The
second (KOI~868.01) has an orbital period of 235.9~d.  Another giant
planet candidate (KOI~1466.01) has $P_K = 281.6$~d and was excluded,
and a fourth (KOI~1552.01) was excluded from our sample because
\emph{Kepler} observed it for only five of the eight quarters.

Following \citet{Mann2012}, we calculated the binomial log likelihood
for a flat log distribution with period and a monotonic radius
distribution in the limit that the transit probability is low:
\begin{equation}
\label{eqn.likely2}
\ln \mathcal{L} = \displaystyle\sum_i^D \left[\ln f + \ln D_i(P_i)\right] - \frac{f}{\ln (P_2/P_1)}\displaystyle\sum_j^{ND} F_j,
\end{equation}
where the orbital period range is $P_1 < P_K < P_2$, the two
summations are over detections and non-detections, respectively,
$D_i(P)$ is the probability of detecting a planet around the $i$th
star,
\begin{equation}
\label{eqn.f}
F_j = \int_{P_1}^{P_2} D_j(P) \, d\ln P,
\end{equation}
and an uninteresting constant is ignored.  To compare with the M2K
results we use $P_1 = 1.7$~d and $P_2 = 245$~d.

We assumed that any transit of a giant planet in front of a late K
dwarf will be detected.  The typical transit depth is $\sim 0.02$,
which is far larger than the noise: The median 3~hr Combined
Differential Photometry Precision for these stars is $1.8 \times
10^{-4}$ and the 99\% value of $6.6\times 10^{-4}$, corresponding to
SNR of $\sim$110 and 30, respectively.  \citet{Fressin2013} found that
the recovery rate of the \emph{Kepler} detection pipeline was nearly
100\% for SNR$>$16.  Thus the detection probability is simply the
geometric factor $R_*/a$, where $a$ is the orbital semimajor axis, and
\begin{equation}
D_j(P) = \left(\frac{4\pi^2 R_*^3}{GM_*}\right)^{1/3}\frac{1 + e \cos \omega}{1-e^2} P_K^{-2/3}
\end{equation}
We marginalized $F$ (Eqn. \ref{eqn.f}) over $e$ and $\omega$ and
adopted a distribution $n(e)$ for eccentricity.  Ignoring terms that
do not depend on $f$, Eqn. \ref{eqn.likely2} becomes
\begin{equation}
\begin{split}
\label{eqn.likelihood2}
\ln \mathcal{L} \approx N_D\ln f - 0.356 f \left[\int_0^1\frac{n(e)de}{1-e^2}\right]\left(\frac{P_2}{\rm 1~d}\right)^{-2/3} \\
\times \frac{\left(P_2/P_1\right)^{2/3}-1}{\ln (P_2/P_1)}\displaystyle \sum_j^{ND}\left(\frac{\rho_j}{\rho_{\odot}}\right)^{-1/3},
\end{split}
\end{equation}
where $N_D$ is the number of detected planets and $\rho$ is the mean
density of the star.  Adopting the function for $n(e)$ in
\citet{Shen2008}, we found that the integral is only weakly dependent
on the parameter $a$ in their distribution, and is $\approx 1.20$ for
$a=4$.  Using a Rayleigh distribution like that for the M2K analysis
gives a similar value (1.08) for the integral.  Because each star can
be explained by more than one stellar model with probability $p$, we
used a weighted mean of $\rho^{-1/3}$ to calculate the likelihood:
\begin{equation}
\langle \rho^{-1/3} \rangle = \displaystyle \sum_i p_i \rho_i^{-1/3} /\displaystyle \sum_i p_i,
\end{equation}
where the summation is restricted to main sequence models, i.e. $\log
g > 4$.  

We compared our analysis with that of \citet{Howard2012} by
calculating $f$ for dwarfs with $4100< T_{\rm eff} < 4600$~K and $4600
< T_{\rm eff} < 5100$~K, and restricting the period range to 0.68~d $<
P_K <$ 50~d.  Our results are 0 and 0.3\% for the respective \teff{}
bins, compared to 0 and $2.7^{+1.0}_{-1.4}$\% from \citet{Howard2012}.
Despite the same restrictions on \teff{}, there are differences in the
samples because we reclassified some K stars as giants (and any
candidate giant planets as stellar companions) and we imposed a
$V$-$J$ color cut which excludes many systems from the hotter \teff{}
bin, whereas \citet{Howard2012} required $K_P < 15$.

\section{Results and Discussion \label{sec.discussion}}

\subsection{Fraction of Stars with Giant Planets \label{subsec.results}}

\begin{figure}
\epsscale{0.8}
\plotone{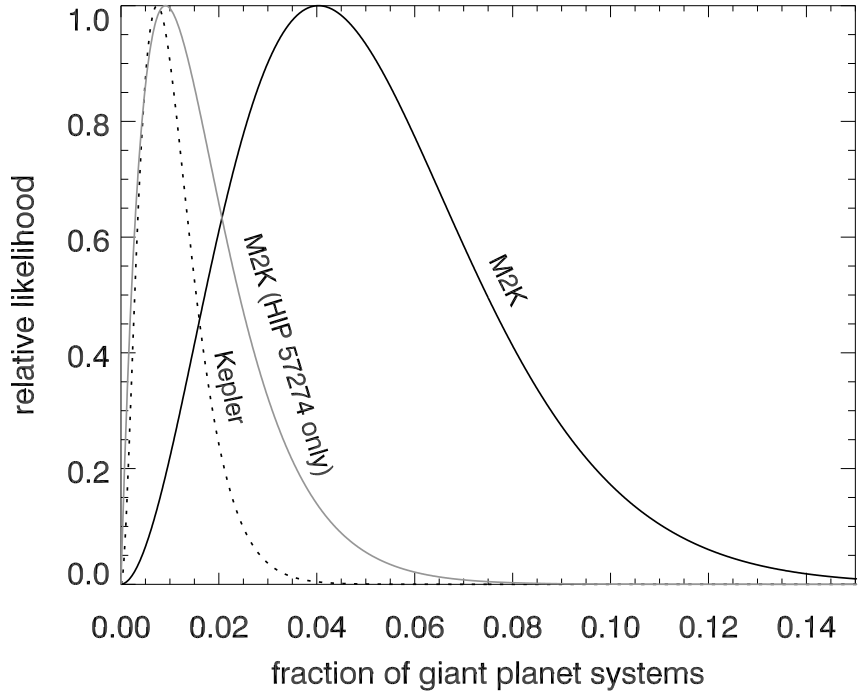}
\caption{Probability distribution of the fraction of 110 M2K stars
  with giant planets with $M_p > 0.3M_J$ and $P_K < 245$~d (solid
  curves).  The curve labeled ``HIP~57274 only'' assumes that such
  planets are ruled out around all but one of the stars: HIP~57274.
  The dashed line is the probability distribution of the fraction of
  \emph{Kepler} late K dwarfs having giant planets with $0.7R_J < R_p
  < 2R_J$ and $P_K < 245$~d. \label{fig.occurrence}}
\end{figure}

Values of $p^1$, the probability that the RV data are consistent with
the presence of a giant planet with $P_K < 245$~d, are reported for
the 110 late K dwarfs in Table 2 (note that $p^0 =
1-p^1$).  Figure \ref{fig.occurrence} shows the relative likelihood
distribution of $f$ constructed from these values using
Eqn. \ref{eqn.likely}. The most probable value of $f$ is 4.0\% and the
uncertainty based on the assumption of asymptotic normality is
$\pm2.3$\%.  The elevated range relative to the rate of actual
detections (1 of 110 stars) is due to the presence of several stars in
our sample with significantly non-zero values of $p^1$.  Eighteen
stars have $p^1 > 0.2$ and four, excluding HIP~57274, have $p^1 > 0.9$
(Table 2).  We are continuing to monitor these stars
(Boyajian et al., in prep.). If we assume that giant planets with $P_K
< 245$~d are ruled out ($p^1 = 0$) for all stars other than HIP~57274
(the ``HIP~57274 only'' case in Fig. \ref{fig.occurrence}), the most
probable value of $f$ becomes $0.92\pm0.75$\%.

Stars may exhibit high RV variation for reasons other than the
presence of giant planets with $P< 245$~d.  Many M2K target stars were
monitored for intervals $\gg 245$~d and our RV data is sensitive to
the presence of planets on wider orbits.  Both Doppler and
\emph{Kepler} surveys find such planets, e.g. HIP~2247 and KOI~1466.01
(see below).  Some stars may have a lower-mass (M dwarf) companion
like that of HIP~38117 (Fig. \ref{fig.hip38117}), but on a wider
orbit.  If the trend in RV produced by such a companion isn't resolved
because of undersampling, it will manifest itself as a high RMS.
Despite our precautionary elimination of stars with Ca II HK emission,
some stars in ours sample may have high intrinsic ``jitter'' from
spots.  Many stars in our sample have only a few Doppler observations
(Fig. \ref{fig.nobshist}), confounding these effects.  Ultimately,
additional observations are required to descriminate between these
possibilities.

The relative likelihood distribution of $f$ for late K dwarfs observed
by \emph{Kepler} is plotted as the dashed line in
Fig. \ref{fig.occurrence}.  The most likely value of $f$ is $0.7 \pm
0.5$\%.  Based on the two distributions, we calculate a 99\%
probability that the \emph{Kepler} value is actually lower than the
M2K value.  Due to the factors discussed above, the M2K value of 4\%
may be an overestimate: There is a closer correspondence (85\% chance
that the \emph{Kepler} estimate is lower) if we rule out giant planets
around all stars other than HIP~57274. \citet{Wright2012} report that
the occurrence of ``hot'' Jupiters around the FGK stars in the
California Planet Search Doppler survey is 1.2\%, compared to 0.4\%
for \emph{Kepler} \citep{Howard2012,Fressin2013}.  \citet{Gaidos2013}
proposed that the difference between the transit and Doppler results
may be due the presence of subgiants in the \emph{Kepler} target
catalog: planets around such stars will be more difficult to detect
and more likely to experience destructive orbital decay.  This
explanation may be less applicable to late K spectral types where the
giant and main sequence branches are more distinguishable, and we
consider orbits with $P_K \gg 10$~d on which orbital decay will be
negligible.  Another explanation for at least some of this difference
is the exclusion of spectroscopic and resolved binaries from the M2K
sample, but not the \emph{Kepler} sample, may enrich for giant
planets, presuming that such binaries are less likely to host planets
for dynamical reasons
\citep[e.g.,][]{Thebault2006,Bonavita2007,Kaib2013}.

\begin{figure}
\epsscale{0.8}
\plotone{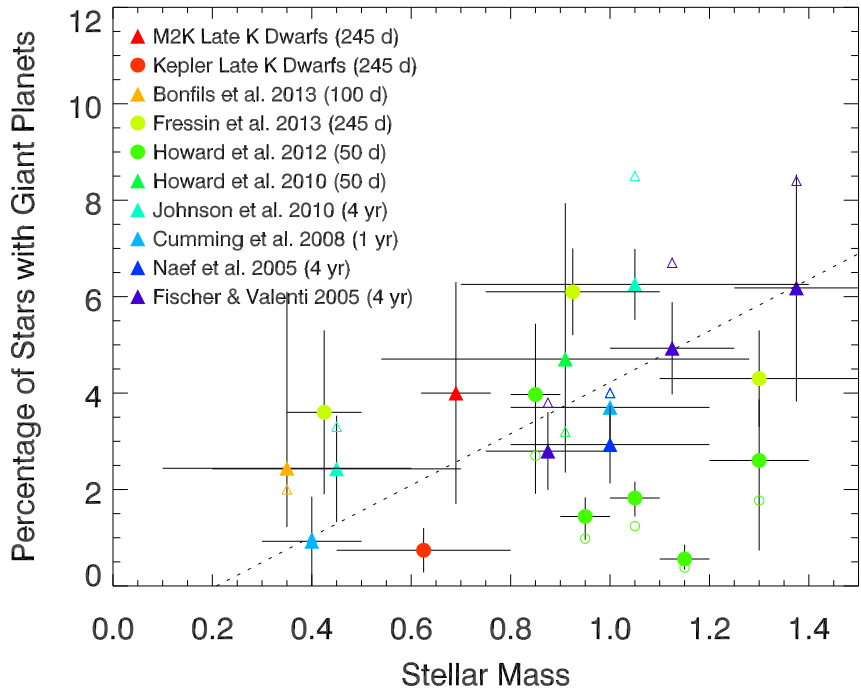}
\caption{Adjusted percentage of stars with giant planets (Saturn mass
  or greater) vs. stellar mass.  Points are color-coded according to
  their source and the symbol indicates whether the estimate is based
  on a RV (triangles) or transit (circles) survey.  Each estimate was
  adjusted by the factor $\ln (P_{max}/{\rm 1.7~d})/\ln (245/1.7)$ to
  acount for different choices of maximum orbital period $P_{max}$
  (see legend).  The unadjusted values are plotted as smaller open
  symbols of the same type and color.  The range of stellar masses is
  in some cases approximate.  The dashed line is a linear
  least-squares fit to the adjusted Doppler
  estimates.\label{fig.compare}}
\end{figure}

We compare our M2K and \emph{Kepler} estimates of $f$ for late K
dwarfs with previous studies for different ranges of stellar mass
(Fig. \ref{fig.compare}).  Our estimates bridge the gap betweeen
solar-type stars
\citep{Fischer2005b,Cumming2008,Johnson2010,Howard2010,Howard2012,Fressin2013}
and M dwarfs
\citep{Naef2005,Cumming2008,Johnson2010,Bonfils2013,Fressin2013}.  We
have adjusted values by the factor $\ln (P_{max}/{\rm 1.7~d})/\ln
(245/1.7)$ to account for differences in the maximum orbital period
$P_{max}$ of each survey, assuming a flat distribution with $\ln P_K$.
(The adjustment is not sensitive to the exact distribution assumed.)
The surveys also differ somewhat in the mass or radius ranges of
objects counted as giant planets.  For example, although our our
\emph{Kepler}-based estimate of 0.7\% for late K dwarfs seems much
lower than those of \citet{Fressin2013} for either GK dwarfs ($6.1\pm
0.9$\%) or M dwarfs ($3.6\pm1.7$\%), these statistics are for $P_K <
418$~d rather than 245~d, and $R_p > 6$\rearth{}, rather than the
8\rearth{} convention adopted here.  Their overall $f$ falls to 4.1\%
for $P < 245$~d, and our $f$ rises to $1.1 \pm 0.6$\% if we include
planets with $R_p > 6$\rearth{}, bringing these two figures closer.
Taken together, these estimates suggest an overall trend, perhaps
linear, of increasing giant planet occurrence with stellar mass, there
is not yet any indication of finer structure.  A linear least-squares
of the adjusted Doppler data yields $f(\%) = -1.11 + 5.33
\,M_*/M_{\odot}$ (dashed line in Fig. \ref{fig.compare}) with weak
significance ($F$-test probability of 0.12).  This compilation also
suggests that the deficit of giant planets around \emph{Kepler} stars
relative to the targets of Doppler surveys \citep{Wright2012} depends
on host star mass (Fig. \ref{fig.compare}), although clearly a more
homogeneous analysis of the collective data sets is needed.

A correlation between giant planets and the metallicity of the host
star has been unambiguously established for solar-type stars
\citep[e.g.,][]{Fischer2005b}, and is strongly supported by the
available evidence for M dwarfs \citep{Neves2013,Mann2013b}.  The
median metallicity of our sample of late K dwarfs is solar ([Fe/H] =
0.004).  The metallicities of HIP~57274 and HIP~2247, the two stars
known to host giant planets in our sample, are 0.08 and 0.24 dex,
consistent with this trend.  The difference in the distributions of
metallicities of stars with $p^1 < 0.1$ (median [Fe/H] of -0.01) and
those with $p^1 > 0.1$ (median [Fe/H] of 0.08) is marginally
significant (K-S probability of 0.06), further supporting a giant
planet-metallicity relation in late K dwarfs.  If SME overestimates
the \teff{} of these stars (Fig. \ref{fig.temp_vs_vj}) , the [Fe/H] is
also overestimated by about 0.1~dex per 100~K.

\subsection{Sensitivity to Parameter Values \label{subsec.sensitivity}}

Our estimates of $f$ may be sensitive to the values of any one of
several parameters we use in our calculations (see Appendix).  These
include the computational resolution $n$ with which $p_i^N$ is
evaluated over ranges of the various orbital parameters, the power-law
indices $\alpha$ and $\beta$ for the assumed mass and period
distributions, the mean value $\bar{e}$ of the Rayleigh-distrbuted
eccentricities, and the RV jitter $\sigma_0$ which is assumed for each
star.  Due to the computational requirements of such studies, we first
considered the effect on two stars, HIP~37798 and HIP~66074, with
number of observations equal to the median ($N=9$) but with the
smallest (0.005) and large (0.82) values of $p^1$, respectively.
Based on the outcome sensitivity of $p^1$ to varying parameter values,
we selectively investigated the effects on our estimates of $f$.

Varying $n$ from 25 to 50 (at rapidly increasing computational cost)
had a negligible effect on $p^1$ for HIP~37798 but decreased the value
for HIP~66074 by about 13\%.  We found that $p^1$ varied little for $n
> 50$.  Thus we re-analyzed 17 stars with $p^1$ values $>0.2$
(excluding HIP~57274) using $n = 50$.  Not all stars were re-analyzed
because of the high computational cost.  These $n = 50$ values are
used in the calculations of $f$ in Section \ref{subsec.results}.
Without the substitution of high-resolution values, the most likely
value of $f$ is $5.1\pm2.7$\%.

We varied the power-law index $\alpha$ of the planet mass distribution
by $\pm0.2$ from its nominal value of -0.31 based on
\citet{Cumming2008} \citep[see also ][]{Howard2012}.  $p^1$ increased
significantly and systematically with more negative values of
$\alpha$, by a factor of 3.5 for HIP~37798 and nearly 1.5 for
HIP~66074.  We found that the most probable value of $f$ changed from
3\% to 6.4\% when we varied $\alpha$ from -0.11 to -0.51.  Also based
on \citet{Cumming2008}, we varied the power-law index $\beta$ of the
orbital period distribution by $\pm0.1$ from its nominal value of
0.26.  We found that the $p^1$ for HIP~37798 was essentially
unchanged, while that of HIP~66074 changed by only $\pm$15\%.  Varying
$\bar{e}$ by $\pm0.1$ from its nominal value of 0.225
\citep{Moorhead2011a}) also had a negligible effect on the $p^1$ of
HIP~37798 and changed that of HIP~66074 only slightly.  Thus our
estimate of $f$ is not not sensitive to the assumed distributions of
orbital period and eccentricity, but does depend on the mass
distribution.  The last occurs because a steeper mass function (more
negative $\alpha$) includes more Saturn-mass planets that could be
hidden on low inclination orbits in our RV data.

Finally, we varied the value of $\sigma_0$ assigned to each star to
account for astrophysical noise and barycenter motion induced by small
planets.  We considered a range of 6-6.75~\mpers{} based on where the
K-S probability that the observed and simulated RV RMS distributions
are within a factor of 0.05 (i.e. 95\% confidence) of the maximum at
$\sigma_0 = 6.3$~\mpers{} (see Sec. \ref{subsec.estimate}).  For both
stars, values of $p^1$ increase significantly if $\sigma_0$ is
decreased from its nominal value, but increased only slightly for
higher $\sigma_0$.  Correspondingly, $f$ increased by a factor of 1.6
for $\sigma_0 = 6$~\mpers{}, and decreased by 0.82 for $\sigma_0 =
6.8$~\mpers{}.  A smaller $\sigma_0$ means that the more RV variation
must be explained by the presence of giant planets, e.g., Saturn-mass
planets with low orbital inclinations.

\subsection{Implications for theory \label{subsec.theory}}

A correlations between the occurrence of giant planets and stellar
metallicity \citep[e.g.,][]{Fischer2005b,Johnson2010} has been
intepreted as supporting the core accretion scenario of giant planet
formation.  In that scenario, growth of a sufficiently massive solid
core leads to the runaway accretion of gas, but only if it occurs
before the gas disk is dissipated in a few million years
\citep{Lissauer2007_PPV}.  In disks of higher metallicity gas, dust
grains can grow, collide, and settle to the mid-plane more rapidly,
thus initiating planet formation at an earlier epoch
\citep{Johnson2012}.  Simulations of rocky planet mass by
\citet{Kokubo2006} produced a linear trend between final planet mass
and initial disk mass surface density.  Thus, disks around
high-metallicity disks should produce larger rocky cores around which
gas could accrete more quickly.

However, a trend with stellar mass, supported by our results, may
require a more complex explanation.  First, the dependence of disk
mass on stellar mass appears to be weak \citep{Williams2011} and
higher disk mass need not translate into higher mass surface density
-- and more massive planets \citep{Kokubo2006} -- if the radial extent
of the disk is larger.  Moreover, Doppler and transit surveys of FGK
stars thoroughly probe orbital semimajor axes to $\lesssim$1~AU;
available radial velocity data suggest a ``jump'' in the population of
giant planets just beyond 1~AU \citep{Wright2009} and set generous
lower limits on their occurrence on much wider orbits
\citep[e.g.,][]{Wittenmyer2011}.  Microlensing surveys suggest that as
many as a third of lensing stars (typically late K and M dwarfs) host
giant planets at 1-5~AU \citep{Mann2010,Cassan2012}.  If giant planet
formation preferentially occurs on these orbits, the correlation with
stellar mass may arise from varying efficiency of inward migration,
rather than formation, of giant planets.

\subsection{On the shoulder of giants \label{subsec.future}}

The coolest giant planet host stars in our M2K and \emph{Kepler}
samples are HIP~57274 and KOI~1176.01 with temperatures of 4640~K and
4625~K.  The only cooler K dwarfs hosting reported giant planets are
WASP-43, HIP~70849, and WASP-80 (Table \ref{tab.planets}), but only
the WASP planets are on close-in orbits.  The effective temperature of
WASP-43, based on the shape of the Balmer H$\alpha$ line, is 4400~K
\citep{Hellier2011}, and this is broadly consistent with the $V-J$
color of 2.4.  However, this star is active and chromospheric emission
may fill in and weaken the H$\alpha$ line, making the temperature
estimate erroneously low.  An analysis of transit light curves coupled
with stellar models suggests $4520\pm120$K instead \citep{Gillon2012}.
The \teff{} assigned to HIP~70849 was based solely on its
luminosity and a theoretical temperature-luminosity relation
\citep{Segransan2011}.  WASP-80 shares spectral characteristics with
both K7 and M0 dwarfs and analyses of a spectrum and infrared
photometry suggests temperatures of $4145\pm100$ and $4020\pm130$K,
respectively \citep{Triaud2013}.

Depending on the properties of WASP-43 and WASP-80, these stars may
bracket a \teff{} range of 4100-4600~K over which giant planets
on close orbits have yet to be found.  This could be a hint of
structure, i.e a gap or ``shoulder'' in the giant planet distribution
with stellar mass, but any conclusion require new surveys.  Giant
planets appear to orbit at wider separations around such stars (e.g.,
HIP~70849b and KOI~868.01), and future space-based astrometric
searches with the \emph{Gaia} mission \citep{deBruijne2012} and
microlensing surveys by \emph{Euclid} \citep{Penny2012} or the
proposed WFIRST observatory \citep{Barry2011} should reveal such
planet populations in detail.

We have used the M2K and \emph{Kepler} surveys to place approximate
constraints on the fraction of late K dwarfs with giant planets, but
the target catalogs are of inadequate size to address the question of
any ``fine structure'' in the distribution of giant planets with
stellar mass.  The Next Generation Transit Survey (www.ngtransits.org)
will monitor 40,000 late G- to early M-type stars to search for
``hot'' Neptunes.  Based on our inferred occurrence rate we expect
there to be $\sim 10$ Jupiters around these target stars, however most
of these will have orbital periods $>10$~d where the detection
efficiency of a ground-based survey at a single site like NGST is low.
The Transiting Exoplanet Survey Satellite will survey $\sim$2.5
million stars to $V = 13$ \citep{Deming2009} and, according to the
TRILEGAL stellar model of the Galaxy \citep{Girardi2012},
approximately 50,000 targets will be late K dwarfs with $4000 < T_{\rm
  eff} < 4800$K.  Monitoring of these should significantly improve the
statistics and allow us to see further.

\clearpage

\acknowledgments

This research was supported by NSF grants AST-09-08406 and NASA grants
NNX10AI90G and NNX11AC33G to EG.  Some of the data presented herein
were obtained at the W.M. Keck Observatory, which is operated as a
scientific partnership among the California Institute of Technology,
the University of California and the National Aeronautics and Space
Administration. The Observatory was made possible by the generous
financial support of the W.M. Keck Foundation.  EG and DF thank the
University of Hawaii and Yale Time Allocation Committees for the
allocations of Keck nights used for this project.  The \emph{Kepler}
mission is funded by the NASA Science Mission Directorate, and data
were obtained from the Mukulski Archive at the Space Telescope Science
Institute, funded by NASA grant NNX09AF08G, and the NASA Exoplanet
Archive at IPAC.

\appendix

\section{Likelihood estimation of planet
  occurrence \label{sec.occurrence}}

In a survey where any giant planet (within the allowed orbital period
range) would be detected, and non-detections unambiguously rule out
planets, the fraction of stars with giant planets $f$ can be
calculated by maximizing the binomial probability distribution for $D$
detections among $N$ systems,
\begin{equation}
P = \frac{N!}{D!(N-D)!}f^{D}(1-f)^{N-D}.
\end{equation}
The first factor can be ignored because it does not depend on $f$,
allowing the problem to be translated into maximimizing a log
likelihood:
\begin{equation}
\label{eqn.likelihood}
\log \mathcal{L} = D \log f + (N-D) \log (1-f).
\end{equation}
However, most of our RV data are ambiguous in that they are neither
detections nor can they rule out all possible giant planets.
Specifically, they can only exclude planets of a certain minimum mass
or minimum inclination with certain combinations of other orbital
parameters.

Equation \ref{eqn.likelihood} can be generalized to $\log L =
\displaystyle\sum_i \log \ell_i(f)$, where $\ell_i(f)$ is the
probability that the RV data of the $i$th star can be explained by a
value of $f$.  Further, the parameter $f$ describes the underlying
probability distribution for the presence or absence of a giant
planet, which in turn generates a model of the RV data.  Using an
empirical Bayes/marginalized likelihood approach this is expressed as
a posterior probability
\begin{equation}
\label{eqn.marglikelihood}
\ell_i(f) = \displaystyle\sum_{N=0}^{1} p(D_i|M_N)q(M_N|f),
\end{equation}
where $p(D_i|M_N)$ is the probability that the $i$th RV data set can
be explained by a model $M_N$ with $N$ planets, $q(M_N|f)$ is the
prior probability of $M_N$ given $f$, and the likelihood is
marginalized over the number of planets.  We seek the value of the
``hyperparameter'' $f$ that maximizes
\begin{equation}
\log \mathcal{L} = \displaystyle \sum_i \log \left[p_i^0(1-f) + p_i^1f \right],
\end{equation}
where $p_i^N = p(D_i|M_N)$ marginalized over all other model
parameters.  Because we expect $f$ to be $\ll 1$, we neglect
multiple-giant planet models.

Assuming Gaussian errors in RV,
\begin{equation}
\label{eqn.prob2}
p_i^n = \exp \left[-\displaystyle\sum_j \frac{\left(v_j-\hat{v}_j^n\right)^2}{2\sigma_j^2}\right]\tilde{p},
\end{equation}
where $v_j$ are the RV measurements, $\hat{v}_j^n$ are the model
values for $n = 0$ or 1 exoplanets, $\sigma_j$ are the errors, and
$\tilde{p}$ represents the product of priors on the model parameters.

The radial velocity model $\hat{v_j}$ of a single planet around a star
depends on six parameters: barycenter velocity $v_0$, $M_p \sin i$,
where $i$ is the inclination, orbital period $P_K$, eccentricity $e$,
argument of periastron $\omega$, and epoch of zero true anomaly $t_0$.
We express this as $\hat{v_j} = v_0 + Kg(t_j)$, where $K$ is the
amplitude of the reflex motion,
\begin{equation}
g(t_j) \equiv \cos \left(\nu_j(t_j) + \omega \right) + e \cos \omega,
\end{equation}
and $\nu_j$ is the true anomaly of the planet at epoch $t_j$.  In the
limit where $M_p \ll M_*$ the reflex amplitude is
\begin{equation}
\label{eqn.semiamp}
K = \frac{M_p \sin i}{M_* \sqrt{1-e^2}}\left(\frac{2\pi G M_*}{P_K}\right)^{1/3}.
\end{equation}
The true anomaly is found by solving for the eccentric and mean
anomalies $\eta$ and $\mu$,
\begin{equation}
\cos \nu = \frac{\cos \eta - e}{1 - e \cos \eta}, 
\end{equation}
\begin{equation}
\mu = \eta - e \sin \eta,
\end{equation}
and
\begin{equation}
\mu = \frac{2\pi}{P}\left(t_j-t_0\right).
\end{equation}
In the single planet model $\hat{v}_j = v_0$, which is the only
parameter in this case.

A precise calculation of $p_i^N$ must marginalize over all possible
parameter values weighted by $\tilde{p}$.  This is computationally
expensive, but if $\sigma \ll K$, Eqn.  \ref{eqn.prob} is very
sensitive to $\hat{v}_j$, and only best-fit parameters will make
significant contributions to $p_i^N$.  The best-fit value of $v_0$ for
a star without a planet is $v_0^* = \langle v_j \rangle$, where
\begin{equation}
  \langle x \rangle \equiv \displaystyle \sum_j x_j \sigma_j^{-2} / \displaystyle \sum_j \sigma_j^{-2}.
\end{equation}
For a star with a planet,
\begin{equation}
v_0^* = \frac{\langle v \rangle \langle g^2 \rangle - \langle vg\rangle \langle g \rangle}{\langle g^2 \rangle - \langle g \rangle^2},
\end{equation}
and the best-fit $K$ is
\begin{equation}
K^* = \frac{\langle vg \rangle - \langle v \rangle \langle g \rangle}{\langle g^2 \rangle - \langle g \rangle^2}.
\end{equation}

Each possible orbit is weighted by a prior for planet mass
distribution and a prior for orbital inclination (the latter is simply
$\sin i$).  However, for a given $K$, $M_*$, $e$ and $P_K$, Eqn.
\ref{eqn.semiamp} inversely relates $M_p$ to a unique value of $\sin
i$.  Thus a marginalization over both parameters collapses to a single
integral over inclination.  For a power-law mass distribution with
index $\alpha$
\begin{equation}
\label{eqn.massweight}
\tilde{p} = C\int_0^{\pi/2} di \int_{M_1}^{M_2} \frac{dM_p}{M_p} \; \left(\frac{M_p}{M_1}\right)^{\alpha} \sin i \ = C \int_0^{\pi/2} di \, \left(\frac{M_p \sin i}{M_1}\right)^{\alpha} (\sin i)^{-\alpha} \cos i
\end{equation}
where the normalization constant $C = -\alpha
\left[1-(M_1/M_2)^{-\alpha}\right]$ and $M_1$ and $M_2$ are the lower
and upper bounds to the mass range.  Equation \ref{eqn.massweight} evaluates to:
\begin{equation}
\label{eqn.massprior}
\tilde{p} = \frac{-\alpha}{1-\alpha}\;\frac{M_p \sin i}{M_1}\;\frac{1-(M_1/M_2)^{1-\alpha}}{1-(M_1/M_2)^{-\alpha}}.
\end{equation}
The lower bound $M_1$ is either $0.3M_J$ (the mass of Saturn) or $M_p
\sin i$, whichever is larger, and $M_2 = 13M_J$, the approximate limit
for deuterium burning in brown dwarfs.  $M_P \sin i$ is uniquely
determined by $K^*$, $M_*$, $e$ and $P_K$.  We adopt $\alpha = -0.31$
based on \citet{Cumming2008}.

Equation \ref{eqn.massprior} is substituted into Eqn. \ref{eqn.prob2}
and marginalized over $\omega \in [0,2\pi]$, $t_0 \in [0,P_K]$, and $e
\in [0,1]$.  The first two are uniformly distributed, and the third is
assumed to be distributed according to a Rayleigh function with a mean
value of 0.225 \citep{Moorhead2011a}.  The only remaining parameter is
orbital period $P_K$.  We marginalize $p_i^N$ over values of $P_K$
drawn from a distribution $P_1 < P_K < P_2$ in a manner that
reproduces a power-law distribution with index $\beta = 0.26$
\citep{Cumming2008}, with $P_1 = 1.7$~d and $P_2 = 245$~d.  For better
sampling of the best-fit values of $P_K$, we iteratively re-calculate
this set of orbital periods using the Keplerian orbital fitting code
RVLIN \citep{Wright2009}, holding other parameters fixed to their
best-fit values, and iterating three time.  We normalize the values of
$p_i^N$ such that $p_i^0 + p_i^1 = 1$, and then evaluate the
likelihood distribution of $f$ using Eqn. \ref{eqn.marglikelihood}.

%\bibliographystyle{/Users/PEBL/Desktop/Bibliography/apj}
%\bibliography{/Users/PEBL/Desktop/Bibliography/apj-jour,/Users/PEBL/Desktop/Bibliography/additional,/Users/PEBL/Desktop/Bibliography/allrefs}

\clearpage

\begin{deluxetable}{lllllll}
  \tabletypesize{\scriptsize} 
%\rotate 
  \tablecaption{Confirmed Giant Planets around Mid- and Late-K-type Dwarf Stars\tablenotemark{a} \label{tab.planets}}
  \tablewidth{0pt}
  \tablehead{\colhead{Planet} & \colhead{Mass ($M_J$)} & \colhead{Period (d)} & \colhead{SpT} & \colhead{$B-V$} & \colhead{\teff{}} & \colhead{References}} 
\startdata
  WASP-80b & 0.55 & 3.068 & K7-M0 & 0.94 & $\sim$4000 & \citet{Triaud2013} \\
  HIP~70849\tablenotemark{b} & $>3$ & $>5$~yr & K7 & 1.42 & 4100\tablenotemark{c} & \citet{Segransan2011} \\
  WASP-43b & 2.03 & 0.813 & K7 & 1.0 & 4400\tablenotemark{d} & \citet{Hellier2011}\\
  HAT-P-20b & 7.25 & 2.88 & K3 & --- & 4619 & \citet{Bakos2011_ApJ742,Torres2012}\\
  HIP~57274c & 0.41\tablenotemark{e} & 32.0 & K4 & 1.11 & 4640 & \citet{Fischer2012}\\
  HIP~57274d & 0.53\tablenotemark{e}  & 431.7 & & & & \\
  WASP-59b & 0.0.86 & 7.92 & K5 & 0.92 & 4650\tablenotemark{h} & \citet{Hebrard2013}\\
  HD~113538b & 0.27\tablenotemark{e}  & 263.3 & K9\tablenotemark{f} & 1.38 & 4685 & \citet{Moutou2011}\\
  HD~113585c & 0.71\tablenotemark{e}  & 1657 & & & & \\
  HIP~2247b & 5.12\tablenotemark{e}  & 655.6 & K4 & 1.14 & 4714 & \citet{Moutou2009}\\
  WASP-10b & 3.06 & 3.09 & K5 & --- & 4735 & \citet{Christian2009,Torres2012}\\
  BD~-08~2823b & 0.33 & 237.6 & K3 & 1.07 & 4746 & \citet{Hebrard2010}\\
  HD~20868b & 1.99 & 380.85 & K3/4 & 1.04 & 4795 & \citet{Moutou2009}\\
  HD~63454b & 0.38\tablenotemark{e}  & 2.82 & K4 & 1.06 & 4840 & \citet{Moutou2005}\\
  Qatar-1b & 1.09 & 1.42 & N/A & 1.06 &  4861 & \citet{Alsubai2011}\\
  HIP~5158b\tablenotemark{g} & 1.44\tablenotemark{e}  & 345.6 & K5 & 1.08 & 4962 & \citet{LoCurto2010}\\
 \enddata
%% Text for table notes should follow after the \enddata but before
%% the \end{deluxetable}. Make sure there is at least one \tablenotemark
%% in the table for each \tablenotetext.
%\tablecomments{This is a table comment.}
 \tablenotetext{a}{$M_p \sin i > 0.3M_J$ and K dwarf hosts with
   spectral subtypes 4 or later in the Exoplanet Catalog
   \citep{Schneider2011} or $1 < B-V < 1.5$ in the Exoplanets Data
   Explorer \citep{Wright2011}}

 \tablenotetext{b}{Imcomplete orbit and parameters are poorly
   constrained.}  

 \tablenotetext{c}{Based on infrared photometry and the
   temperature-luminosity relation of \citet{Baraffe1998}}
 
 \tablenotetext{d}{\teff{} based on H$\alpha$ and may need to be revised
   upward based on new mass estimate \citep{Gillon2012}}
 
 \tablenotetext{e}{$M_P \sin i$} 

 \tablenotetext{f}{Stellar parameters are problematic:
   \citet{Gray2006} assigns it the unrecognized spectral type K9, and
   its $B-V$ and $V-J$ colors suggest a star at the K-M spectral type
   boundary.  \citet{Gray2006} also assign it a ``k'' to indicate
   interstellar absorption features, seemingly inconsistent for a star
   only 16~pc away.  \citet{Moutou2011} and \citet{Bailer-Jones2011}
   assign \teff{} of 4685~K and 4625~K based on spectra and photometry,
   respectively.  To reconcile the \teff{} and colors,
   \citet{Bailer-Jones2011} estimate $\sim$1~mag of extinction, also
   inconsistent with its proximity.}
 
 \tablenotetext{g}{This system also includes HIP~5185c, which may be a
   brown dwarf \citep{Feroz2011}.}  \tablenotetext{h}{\teff{} was
   estimated by the null dependence of abundance on excitation
   potential.}
\end{deluxetable}

\begin{deluxetable}{lrrrrrrrr}
\tabletypesize{\scriptsize}
\tablecaption{Stars Included in the Analysis}
\label{tab.stars}
\tablewidth{0pt}

\tablehead{\colhead{Name} & \colhead{{\it V-J}} & \colhead{$T_{\rm eff}$ (K)} & \colhead{[Fe/H]} & \colhead{$S$} & \colhead{$M$ ($M_{\odot}$)} & \colhead{$N_{\rm obs}$} & \colhead{RMS (m s$^{-1}$)} & \colhead{$p^1$}}
\startdata
    HIP 1078 & 2.13 & 4426$^{b}$ & --- & 0.49 & 0.75 &   4 &   2.56       & 0.005 \\
    HIP 1532 & 2.52 & 3956$^{a}$ & -0.37 & 0.84 & 0.64 &  24 &   4.28       & 0.011 \\
    HIP 2247 & 2.06 & 4680$^{ }$ &  0.24 & 0.62 & 0.77 &  12 &   0.00       & 0.000 \\
    HIP 3418 & 2.16 & 4546$^{ }$ &  0.02 & 0.50 & 0.73 &   5 &   1.54       & 0.004 \\
    HIP 4353 & 2.20 & 4587$^{ }$ &  0.17 & 0.70 & 0.77 &   6 &  12.89       & 0.961 \\
    HIP 4454 & 1.99 & 4671$^{ }$ & -0.55 & 0.35 & 0.71 &   5 &   3.62       & 0.004 \\
    HIP 4845 & 2.58 & 4170$^{ }$ & -0.19 & 1.67 & 0.63 &  33 &   3.55       & 0.005 \\
    HIP 5247 & 2.51 & 4260$^{ }$ & -0.22 & 0.63 & 0.66 &  12 &   7.76       & 0.052 \\
    HIP 5663 & 2.34 & 4357$^{ }$ & -0.05 & 1.04 & 0.68 &  24 &   0.00       & 0.000 \\
    HIP 6344 & 2.32 & 4383$^{ }$ & -0.04 & 0.99 & 0.68 &   5 &   1.71       & 0.007 \\
    HIP 9788 & 2.15 & 4444$^{ }$ & -0.37 & 0.73 & 0.67 &  17 &   4.53       & 0.008 \\
   HIP 10337 & 2.54 & 4019$^{a}$ &  0.12 & 1.58 & 0.66 &  22 &   7.74       & 0.019 \\
   HIP 10416 & 1.88 & 4743$^{ }$ &  0.14 & 0.71 & 0.76 &  11 &  10.43       & 0.721 \\
   HIP 11000 & 1.92 & 4703$^{ }$ &  0.20 & 0.76 & 0.73 &  13 &  10.54       & 0.195 \\
   HIP 12493 & 2.25 & 4350$^{ }$ & -0.29 & 0.66 & 0.68 &   6 &   4.50       & 0.032 \\
   HIP 13375 & 2.50 & 4110$^{b}$ &  0.00 & 0.59 & 0.56 &   9 &   3.68       & 0.010 \\
   HIP 14729 & 2.15 & 4579$^{ }$ &  0.17 & 1.00 & 0.70 &  10 &   5.76$^{c}$ & 0.007 \\
   HIP 15095 & 2.28 & 4265$^{ }$ & -0.21 & 1.28 & 0.68 &  10 &   6.82       & 0.071 \\
   HIP 15563 & 2.06 & 4718$^{ }$ &  0.16 & 1.04 & 0.72 &  12 &  10.74       & 0.332 \\
   HIP 15673 & 1.93 & 4654$^{ }$ & -0.46 & 0.50 & 0.69 &   6 &   3.12       & 0.007 \\
1234-00069-1 & 1.90 & 4522$^{a}$ & --- & 0.42 & 0.75 &   4 &   1.30       & 0.023 \\
   HIP 17346 & 2.06 & 4613$^{ }$ &  0.07 & 0.59 & 0.74 &   6 &   4.19       & 0.012 \\
   HIP 17496 & 2.15 & 4489$^{ }$ & -0.04 & 0.62 & 0.72 &   5 &   2.53       & 0.008 \\
   HIP 19165 & 2.24 & 4412$^{ }$ & -0.22 & 0.64 & 0.65 &  31 &   5.16       & 0.010 \\
   HIP 19981 & 2.25 & 4605$^{ }$ &  0.27 & 0.70 & 0.77 &   5 &   4.60       & 0.017 \\
   HIP 20359 & 2.14 & 4547$^{ }$ & -0.04 & 0.58 & 0.75 &   5 &   4.47       & 0.011 \\
   HIP 25220 & 2.04 & 4613$^{ }$ &  0.05 & 0.89 & 0.71 &  10 &   7.92       & 0.049 \\
   HIP 26196 & 2.34 & 4245$^{b}$ & --- & 0.76 & 0.74 &   8 &   4.26$^{d}$ & 0.012 \\
4356-01014-1 & 1.88 & 4737$^{ }$ & -0.17 & 0.46 & 0.77 &   8 &   2.69       & 0.003 \\
   HIP 29548 & 2.32 & 4528$^{ }$ & -0.10 & 0.59 & 0.69 &  20 &   3.79       & 0.006 \\
   HIP 30112 & 2.36 & 4168$^{a}$ & --- & 1.46 & 0.72 &  25 &   7.45       & 0.040 \\
   HIP 30979 & 2.02 & 4627$^{ }$ &  0.23 & 0.60 & 0.77 &  12 &   5.41       & 0.005 \\
3388-01009-1 & 2.36 & 4220$^{b}$ & --- & 1.38 & 0.68 &  10 &  64.66       & 0.500 \\
   HIP 32769 & 2.24 & 4420$^{ }$ & -0.05 & 0.68 & 0.71 &  16 &   4.86       & 0.005 \\
   HIP 32919 & 2.33 & 4382$^{ }$ & -0.01 & 0.86 & 0.70 &  18 &   5.29       & 0.006 \\
1352-01588-1 & 2.31 & 4270$^{b}$ & --- & 0.00 & 0.69 &   7 &  18.27       & 0.918 \\
0748-01711-1 & 2.50 & 4064$^{a}$ & --- & 0.16 & 0.66 &  25 &  42.28$^{c}$ & 0.500 \\
   HIP 36551 & 2.09 & 4501$^{ }$ & -0.30 & 0.57 & 0.70 &  11 &   3.86       & 0.007 \\
   HIP 37798 & 2.54 & 4082$^{b}$ & --- & 1.42 & 0.70 &   9 &   3.12       & 0.005 \\
   HIP 38969 & 2.13 & 4761$^{ }$ &  0.26 & 0.40 & 0.81 &  10 &  10.96$^{c}$ & 0.817 \\
   HIP 40375 & 2.18 & 4463$^{ }$ &  0.03 & 0.96 & 0.71 &  33 &   6.67       & 0.009 \\
   HIP 40671 & 2.05 & 4612$^{ }$ &  0.06 & 0.46 & 0.74 &   6 &   1.95       & 0.008 \\
   HIP 40910 & 2.41 & 4119$^{a}$ & -0.06 & 1.41 & 0.68 &  21 &   9.47       & 0.665 \\
   HIP 41130 & 2.26 & 4410$^{ }$ & -0.10 & 1.10 & 0.72 &  18 &   8.46       & 0.603 \\
   HIP 41443 & 2.08 & 4613$^{ }$ &  0.01 & 0.83 & 0.74 &   9 &   6.10       & 0.014 \\
   HIP 42567 & 1.94 & 4648$^{ }$ &  0.09 & 0.71 & 0.76 &   6 &   6.00       & 0.037 \\
   HIP 43534 & 2.56 & 4100$^{a}$ & -0.13 & 1.49 & 0.64 &  14 &   6.88       & 0.006 \\
   HIP 43667 & 2.12 & 4554$^{ }$ &  0.01 & 0.51 & 0.72 &   9 &   5.24       & 0.020 \\
   HIP 44072 & 2.16 & 4347$^{ }$ & -0.42 & 0.52 & 0.71 &   5 &   3.96       & 0.020 \\
   HIP 45042 & 2.35 & 4476$^{ }$ &  0.17 & 1.27 & 0.73 &   5 &   9.94       & 0.095 \\
1955-00658-1 & 2.52 & 3991$^{a}$ &  0.21 & 1.48 & 0.66 &   4 &   6.05       & 0.055 \\
   HIP 45839 & 2.12 & 4590$^{ }$ &  0.05 & 0.55 & 0.72 &   5 &  12.40       & 0.865 \\
   HIP 46343 & 2.20 & 4529$^{ }$ &  0.03 & 1.03 & 0.70 &  10 &   4.35       & 0.008 \\
   HIP 46417 & 2.28 & 4475$^{ }$ & -0.06 & 0.69 & 0.71 &   9 &   7.19       & 0.018 \\
   HIP 47201 & 2.37 & 4122$^{a}$ &  0.03 & 0.98 & 0.69 &   7 &   4.91       & 0.045 \\
   HIP 48139 & 2.09 & 4548$^{a}$ &  0.22 & 0.45 & 0.77 &   6 &   2.34       & 0.004 \\
   HIP 48411 & 2.14 & 4505$^{ }$ &  0.20 & 0.78 & 0.73 &   4 &   8.32       & 0.057 \\
   HIP 48740 & 2.22 & 4577$^{ }$ &  0.02 & 1.10 & 0.72 &   8 &   7.38       & 0.015 \\
   HIP 50960 & 2.40 & 4187$^{a}$ & -0.06 & 1.51 & 0.65 &   8 &   8.22       & 0.023 \\
   HIP 51443 & 2.16 & 4505$^{ }$ & -0.05 & 1.09 & 0.71 &  19 &   8.56       & 0.447 \\
   HIP 53327 & 2.29 & 4390$^{ }$ & -0.79 & 0.37 & 0.66 &   4 &   5.13       & 0.018 \\
   HIP 54459 & 2.26 & 4469$^{ }$ & -0.52 & 0.37 & 0.68 &  13 &   8.32       & 0.028 \\
   HIP 54651 & 2.11 & 4395$^{ }$ & -0.89 & 0.33 & 0.66 &   6 &   1.63       & 0.002 \\
   HIP 54810 & 2.09 & 4256$^{a}$ &  0.03 & 1.04 & 0.70 &   5 &   6.18       & 0.061 \\
   HIP 55507 & 2.42 & 4104$^{a}$ & -0.05 & 0.90 & 0.69 &  22 &   6.37$^{c}$ & 0.013 \\
   HIP 56630 & 2.67 & 3960$^{a}$ & -0.01 & 1.37 & 0.68 &   9 &   7.29       & 0.045 \\
   HIP 57274 & 2.00 & 4641$^{ }$ &  0.08 & 0.39 & 0.76 & 120 &   0.00       & 1.000 \\
   HIP 57493 & 2.38 & 4168$^{a}$ &  0.06 & 1.16 & 0.71 &  15 &   5.48       & 0.009 \\
   HIP 59496 & 2.41 & 4122$^{a}$ & -0.01 & 1.42 & 0.69 &   7 &   9.67       & 0.203 \\
   HIP 60633 & 2.08 & 4724$^{ }$ &  0.25 & 0.33 & 0.76 &  26 &  12.92       & 1.000 \\
   HIP 62406 & 2.38 & 4168$^{a}$ &  0.31 & 1.10 & 0.68 &  39 &   8.07       & 0.118 \\
   HIP 62847 & 1.97 & 4726$^{ }$ &  0.05 & 0.66 & 0.81 &  34 &   6.03       & 0.012 \\
   HIP 63894 & 2.20 & 4361$^{b}$ & --- & 0.81 & 0.69 &   8 &   6.47       & 0.029 \\
   HIP 64048 & 2.17 & 4615$^{ }$ &  0.08 & 0.71 & 0.71 &  14 &   8.63       & 0.029 \\
   HIP 64262 & 2.03 & 4639$^{ }$ & -0.25 & 0.59 & 0.70 &   9 &   8.13       & 0.047 \\
   HIP 66074 & 2.32 & 4460$^{ }$ &  0.23 & 1.17 & 0.73 &   9 &  11.06       & 0.818 \\
   HIP 66222 & 2.71 & 3861$^{a}$ & -0.11 & 1.34 & 0.68 &  13 &   6.54       & 0.017 \\
   HIP 66283 & 1.84 & 4880$^{ }$ &  0.18 & 0.00 & 0.84 &   7 &  10.71       & 0.256 \\
   HIP 66840 & 2.47 & 4019$^{a}$ & --- & 0.00 & 0.69 &   5 &   1.96       & 0.008 \\
   HIP 67842 & 2.76 & 3919$^{b}$ & --- & 1.23 & 0.64 &   6 &   5.18       & 0.016 \\
   HIP 73427 & 2.40 & 4256$^{a}$ & -0.02 & 1.26 & 0.73 &  11 &  13.32       & 0.949 \\
   HIP 75672 & 2.45 & 4083$^{a}$ & --- & 0.00 & 0.69 &   4 &   2.67       & 0.021 \\
   HIP 77908 & 2.27 & 4612$^{ }$ &  0.28 & 0.98 & 0.74 &   5 &   2.50       & 0.008 \\
   HIP 78184 & 2.47 & 4168$^{a}$ & --- & 0.00 & 0.68 &   4 &   7.08       & 0.004 \\
   HIP 78999 & 2.05 & 4504$^{ }$ &  0.24 & 0.00 & 0.73 &   4 &   6.30       & 0.029 \\
   HIP 79698 & 2.31 & 4256$^{a}$ &  0.26 & 0.91 & 0.74 &   7 &   3.51       & 0.004 \\
   HIP 87464 & 2.17 & 4417$^{ }$ & -0.27 & 0.44 & 0.69 &   8 &   8.76       & 0.351 \\
   HIP 89087 & 2.40 & 4118$^{a}$ & -0.01 & 1.38 & 0.68 &   4 &   3.48       & 0.001 \\
   HIP 93871 & 2.06 & 4453$^{a}$ & -0.47 & 0.28 & 0.71 &   4 &   3.38       & 0.001 \\
   HIP 97051 & 2.34 & 4320$^{ }$ & -0.19 & 1.10 & 0.62 &  19 &   9.10       & 0.183 \\
   HIP 99205 & 2.15 & 4397$^{ }$ & -0.18 & 0.85 & 0.64 &   6 &   6.32       & 0.021 \\
   HIP 99332 & 2.15 & 4400$^{a}$ &  0.30 & 0.73 & 0.76 &   8 &   5.82       & 0.033 \\
  HIP 101262 & 2.06 & 4774$^{ }$ &  0.19 & 1.00 & 0.73 &   4 &   9.06       & 0.146 \\
  HIP 102332 & 2.39 & 4202$^{b}$ & --- & 0.75 & 0.66 &   4 &   5.56       & 0.017 \\
  HIP 103650 & 2.10 & 4603$^{ }$ & -0.04 & 0.75 & 0.71 &   9 &   5.22       & 0.020 \\
  HIP 104092 & 2.23 & 4455$^{ }$ &  0.07 & 0.54 & 0.72 &   5 &   6.66       & 0.057 \\
  HIP 105341 & 2.55 & 4299$^{ }$ & -0.05 & 1.38 & 0.68 &   5 &   8.68       & 0.087 \\
  HIP 109980 & 2.14 & 4580$^{ }$ &  0.00 & 0.95 & 0.71 &  56 &   6.73       & 0.024 \\
  HIP 110774 & 2.14 & 4471$^{ }$ & -0.12 & 0.87 & 0.67 &   4 &   1.91       & 0.000 \\
3995-01436-1 & 2.16 & 4634$^{ }$ &  0.08 & 0.89 & 0.71 &   9 &   7.20       & 0.051 \\
  HIP 112496 & 2.00 & 4778$^{ }$ &  0.04 & 0.65 & 0.75 &   5 &   7.92       & 0.059 \\
  HIP 112918 & 2.26 & 4434$^{ }$ &  0.02 & 1.20 & 0.67 &   4 &   2.80       & 0.018 \\
  HIP 114156 & 2.45 & 4322$^{ }$ & -0.02 & 1.43 & 0.66 &   4 &   3.41       & 0.025 \\
  HIP 115004 & 2.22 & 4684$^{ }$ & -0.11 & 0.83 & 0.82 &  21 &  40.79$^{c}$ & 0.500 \\
  HIP 117197 & 2.56 & 4153$^{ }$ & -0.38 & 1.05 & 0.61 &  19 &   3.15       & 0.002 \\
  HIP 117492 & 2.10 & 4679$^{ }$ &  0.08 & 0.98 & 0.73 &   5 &   5.09       & 0.029 \\
  HIP 117559 & 2.25 & 4581$^{ }$ &  0.07 & 1.28 & 0.71 &  41 &   9.82       & 0.447 \\
  HIP 117946 & 1.95 & 4863$^{ }$ &  0.03 & 0.85 & 0.77 &  14 &   9.00       & 0.143 \\
  HIP 118261 & 1.90 & 4662$^{ }$ & -0.04 & 0.65 & 0.74 &  34 &   7.28       & 0.025 \\
  HIP 118310 & 2.24 & 4326$^{b}$ & --- & 0.71 & 0.80 &   4 &   0.00       & 0.000 \\
\enddata
\tablenotetext{a}{based on medium-resolution spectroscopy}
\tablenotetext{b}{based on $V-J$ color}
\tablenotetext{c}{a linear trend removed}
\tablenotetext{d}{a parabolic trend removed}
\end{deluxetable}

\end{document}